\documentclass[10pt,a4paper,notitlepage]{article}
\newcommand{\be}{\begin{equation}}
\newcommand{\ee}{\end{equation}}
\usepackage{german}
\usepackage{graphicx}
\usepackage{amsfonts} 
\RequirePackage[latin1]{inputenc}
\usepackage[german]{varioref}

\begin{document}
{\huge On the relation between the second law of thermodynamics and classical and quantum mechanics}

\medskip
Barbara Drossel, Institut f\"ur Festk\"orperphysik, Technische Universit\"at Darmstadt, Hochschulstr. 6, 64289 Darmstadt, Germany.

\medskip
{Preliminary version of a contribution to appear in:  B. Falkenburg and M. Morrison (eds.), \textit{Why more is different}
  (Springer Verlag, 2014)}

\section{Introduction}
This article is devoted to the relation between the second law of thermodynamics, which applies to closed macroscopic systems consisting of an extremely large number of particles, such as liquids or gases, and classical or quantum mechanics, which are theories that describe systems of interacting particles on a microscopic level. The second law of thermodynamics states that in a closed system entropy increases until it reaches a maximum. Then the system has reached equilibrium. Entropy is a property of a ``macrostate'', which is characterized by measurable variables such as density $\varrho(\vec x)$ or magnetization $M(\vec x)$, which are sums or averages over many particles. The entropy is assumed to be proportional to the logarithm of the number of different ``microstates'' that correspond to such a macrostate. Based on the concept of microstates, the second law of thermodynamics (and all other relations of thermodynamics) can be obtained from statistical mechanics with its basic axiom that in a closed system in equilibrium all microstates occur with equal probability. This means that transition probabilities between microstates are such that in the long run no state is preferred.

Now, the microscopic description of a many-particle system in terms of classical or quantum mechanics differs in two fundamental ways from the statistical mechanics description, which entails the second law of thermodynamics. First, classical mechanics and quantum mechanics are deterministic theories. Given the initial state of a system, these theories determine its future time evolution. In contrast, statistical mechanics is a stochastic theory, with probability being an important concept. Second, classical mechanics and quantum mechanics are time reversible. If a given trajectory is a solution of Newton's laws, the time inverted trajectory is also a solution, because Newton's laws do not change under time reversal, due to the second derivate with respect to time. Similarly, if a wave fuction $\psi(\vec x, t)$ is a solution of the Schr\"odinger equation, its complex conjugate $\psi^*(\vec x,t)$ is a solution of the time reversed Schr\"odinger equation, giving exactly the same physical properties, since observables depend only on the absolute value of the wave function. In contrast, the second law of thermodynamics makes a fundamental distinction between the two directions of time. The entropy increase occurs only in the forward time direction. 

Because of these two fundamental differences, the question arises if and how the second law of thermodynamics (and statistical mechanics in general) can be derived from classical or quantum mechanics. Many textbook authors assume that in principle a macroscopic system is on the microscopic level fully and correctly described by deterministic, time-reversible laws. Of course, if such a  microscopic description is complete, it must somehow contain all the properties that are perceived in thermodynamic systems that consist of the order of $10^{23}$ particles. Consequently, the irreversible character of the second law is ascribed by these textbook authors to our inability to obtain knowledge of the precise microscopic state of the system, combined with special initial conditions for the macroscopic, observable quantities of the system. This is the so-called ignorance interpretation of probability. Our inability to know the microstate of the system and to thus predict its future evolution is aggravated by the fact that no system is fully isolated from the rest of the world. This means that the future evolution of a system is influenced by its environment. In order to see the deterministic character of a system's time evolution, one would need to know the state of its environment, which in turn depends on still another environment, etc. Since it is impossible to include in our calculations such a wider environment, which might consist of the entire universe, some textbook authors argue that we have no choice but to describe our system by using the concept of probabilities. 

At this point it becomes clear that the belief that the time evolution of a many-particle system is deterministic on the microscopic level is a metaphysical belief. It cannot, even in principle be shown to be correct. It is a starting assumption on which the subsequent considerations are based, and it is not the result of scientific observations. In order to assess how reasonable this basic assumption is, one has to explore its logical consequences. In fact, the debate on how to relate the probabilities of statistical mechanics to a microscopic view is as old as the theory itself, starting from the Boltzmann-Zermelo debate and continuing until today, see for instance the contribution by Jos Uffink in (Beisbart and Hartmann, 2011).

In the following, I will argue that the second law of thermodynamics cannot be derived from deterministic time-reversible theories such as classical or quantum mechanics. This means that even simple macroscopic equilibrium systems such as gases, crystals, or liquids, cannot be fully explained in terms of their constituent particles, or, to use the wording of the title of this book, that ``more is different''.  In the next section, I will challenge the basic assumption of many textbook authors that classical or quantum mechanics can provide an accurate, comprehensive microscopic description of a thermodynamic system. Then, I will show that all so-called ``derivations'' of the basic concepts of statistical mechanics and in particular of the  second law of thermodynamics make - often unadmittedly - assumptions that go beyond classical or quantum mechanics. 

\section{The mistaken idea of infinite precision}

In classical mechanics, the state of a system can be represented by a point in phase space. The phase space of a system of $N$ particles has $6N$ dimensions, which represent the positions and momenta of all particles. Starting from an initial state, Newton's laws, in the form of Hamilton's equations, prescribe the future evolution of the system. If the state of the system is represented by a point in phase space, its time evolution is represented by a trajectory in phase space. However, this idea of a deterministic time evolution represented by a trajectory in phase space can only be upheld within the framework of classical mechanics if a point in phase space has infinite precision. If the state of a system had only a finite precision, its future time evolution would no more be fixed by the initial state, combined with Hamilton's equations. Instead, many different future time evolutions would be compatible with the initial state. In practice, it is impossible to know, prepare, or measure the state of a system with infinite precision. This would require a brain or another computing device that can store an infinite number of bits, which does not exist in a finite universe. Here, we see again that the belief that classical mechanics can provide a valid microscopic description of a thermodynamic system is a metaphysical belief. 

In quantum mechanics, the state of a system is represented by a wave function, and its time evolution by the Schr\"odinger equation. Now, in order for the Schr\"odinger equation to fully predict the future evolution of a quantum mechanical system, not just a point must be specified with inifinite precision, but  a complex-valued function of $3N$ variables, because the wave function is complex and depends on the positions of all particles. Even the ``simple'' enterprise of calculating the ground state wave function of a many-electron system fails completely for more than 1000 particles, says Walter Kohn, the father of density functional theory, in his Nobel lecture, quoting his teacher Van Vleck: ``The many-electron wave function is not a legitimate scientific concept for $N > 1000$ ... because the wave function can neither be calculated nor recorded with sufficient accuracy'' (Kohn, 1999). This illustrates again that the idea that the state of a system has infinite precision is a metaphysical assumption. 

History and philosophy of science have taught us that the idealizations that are contained in the theories of physics should not be taken as a faithful and perfect reflection of reality. In classical mechanics, these idealizations include certain concepts of space and time, in addition to a deterministic worldview. The belief that classical mechanics is a faithful and perfect reflection of physical reality was shattered 100 years ago, when the theory of relativity showed it to be an approximation that is pretty good when velocities are far below the velocity of light and when gravitational fields are weak enough so that the curvature of space cannot be perceived. Furthermore, the advent of quantum mechanics made it clear that classical mechanics is not valid on atomic length scales. In particular, the uncertainty principle states that a point in phase space cannot have infinite precision. This has drastic consequences because the discovery of chaos demonstrated that in many systems a limited precision of the initial state leads to a complete uncertainty of its future time evolution beyond a short time horizon.

In quantum mechanics, we know that the Schr\"odinger equation is a good description of a system only when relativistic effects and radiation effects can be neglected. In addition to these reasons to consider wave functions and the Schr\"odinger equation only as an approximate description of reality, quantum mechanics poses a far more fundamental problem. It has itself a part that is inherently stochastic and irreversible, namely the quantum measurement. With respect to the outcome of a measurement, only probabilities can be given. Furthermore, a measurement process, which includes irreversible changes in the macroscopic measurement device, does not occur backwards in time, thus making a distinction between past and future. Landau and Lifschitz suggest in their textbook on statistical physics that the irreversibility of the measurement process is related to the irreversibility of the second law of thermodynamics. Despite claims made to the contrary by some scientists, the measurement process has not been satisfactorily explained in terms of the deterministic evolution of a many-particle wave function according to the Schr\"odinger equation (Schlosshauer 2004). 

All these well known limitations of classical mechanics and quantum mechanics support my argument that the idea of inifinite precision, which follows from these theories if they are an exact reflection of reality, is a very questionable concept. The belief that the time evolution of a thermodynamic system is deterministic and reversible on the microscopic level is a metaphysical assumption that presents a variety of problems. As soon as the metaphysical assumption of infinite precision is abandoned, the present state of a system, combined with the microscopic laws of classical or quantum mechanics, does not fully specify the future time evolution. Therefore, additional laws are needed that specify which types of time evolutions out of the possible ones are taken by the system. 

I will therefore join those textbook authors that maintain that statistical mechanics is a field of physics in its own right, which is not contained in deterministic, time-reversible microscopic theories. Statistical mechanics has its own axiom, namely the axiom of ``equal a priori probabilities''. This axiom states that in a closed system in equilibrium all microstates occur with equal probability. From this axiom, all of statistical mechanics can be derived. 

In the next section, I will show that even those scientists who ``derive'' the law of equal a priori probabilities and the second law of thermodynamics from classical mechanics, make in fact assumptions that go beyond classical mechanics. Then, we will briefly look at other approaches to the relation between classical and statistical mechanics, which openly employ additional assumptions and thus admit that statistical mechanics cannot be shown to be contained in classical mechanics. In Section 3, we will then  take a short survey of the different possible ways to relate statistical mechanics to quantum mechanics, encountering similar and even worse challenges than for classical mechanics. 

\section{From classical mechanics to statistical mechanics}
\subsection{The standard argument}

As already mentioned, every ``derivation'' of statistical mechanics from classical mechanics starts from the metaphysical assumption that classical mechanics can provide a complete and deterministic description of a $10^{23}$-particle system. An important part of this assumption is the idea that the state of a system has infinite precision. We will now start from this assumption and develop the usual arguments that can be found in many textbooks. We will see that in fact the deterministic assumption is not followed through to the end, but that probabilistic assumptions, which are foreign to classical mechanics, creep in, often without being perceived as such.

An important concept when relating classical mechanics to statistical mechanics is ``quasi-ergodicity'': The concept of quasi-ergodicity means that a ``typical'' trajectory in phase space comes arbitrarily close to every point on the energy shell if one waits long enough. ``Typical'' are all trajectories with exception of a few special trajectories, the initial points of which have a measure zero in the energy shell. Unstable periodic orbits belong to this special class of trajectories. The ``energy shell'' is the subspace of phase space that has the energy of the initial state. Due to the hamiltonian character of time evolution, energy is conserved. Due to the strongly chaotic character, there are no other conserved quantities and supposedly no islands of regular dynamics. If we lay a grid with some cell size $\epsilon^{6N-1}$ over the energy shell, a typical trajectory will have visited all cells of the grid after a time that depends on $\epsilon$. Since there is no reason to prefer any part of the energy shell, all  parts will be visited on an average equally often. This is one way of stating that all microstates of the system occur with equal probability. 

It is also instructive to consider the time evolution of an ensemble of initial states, all of which lie in a small compact volume in phase space. With time, the volume becomes deformed according to Liouville's equation, but its size does not change. Due to the stretching and folding process of chaotic motion, finer and finer filaments of the volume will penetrate to more and more cells of our phase space grid. After some time, all cells will contain a small part of the original volume in the form of  very fine filaments. If the size of the initial droplet presents the precision of our knowledge of the initial state, we cannot say at all in which cell the system will be after this time. It can be anywhere with equal probability. This leads again to the basic axiom of statistical mechanics. 

In order to justify the second law of thermodynamics, an additional consideration is needed: Most cells in the energy shell correspond to the macrostate with the largest entropy. This is a consequence of the huge dimension of phase space. Therefore, the argument goes, even if one starts from a low-entropy initial state, after a short time the system will reach  ``typical'' cells in phase space, which have maximum entropy. 

\subsection{The problems with the standard argument}

There are two important problems with the argument outlined in the previous subsection, both of which are closely tied to the underlying idea that the state of a system corresponds to a point in phase space.

First, as pointed out by several textbook authors, the concept of quasiergodicity poses problems, because the time required to visit every cell in phase space is incredibly much larger than the lifetime of the universe. During the short time of an experiment, where we observe a system to reach equilibrium, only a vanishingly small fraction of phase space can be visited by the system. Therefore, the fundamental theorem of statistical mechanics, i.e., the theorem of equal a priory probabilities, cannot be derived from classical mechanics in this way. Fortunately, quasiergodicity is not required for justifying the second law of thermodynamics, since we only need to argue that each cell in the energy shell is not far from maximum entropy cells, which are by far the most numerous types of cells. 

However, when claiming that a trajectory will after some time ``most likely'' or ``virtually certainly'' be in cells with maximum entropy, one makes a probabilistic argument, which cannot be justified based on classical mechanics alone, and this is the second problem with the standard derivation sketched in the previous subsection. Stated somewhat differently and in more detail, the probabilistic argument must be understood as follows: Given an initial state of the system, consider all future time evolutions that are compatible with this initial state within the given precision. Assuming that all these time evolutions happen with the same probability, and given the fact that the vast majority of these time evolutions show an increase in entropy towards equilibrium, the system will show an increase in entropy and approach equilibrium. Without the possibility of different time evolutions, probabilistic statements make no sense. A strictly deterministic world with infinite-precision phase space points leaves no freedom to ``choose'' the most probably time evolution, because the initial state fully contains the future time evolution. 

A time evolution that does not agree with the ``most probable'' behavior is also compatible with classical mechanics, as one can conclude from the fact that by going backwards in time a system will arrive at the initial state at which it was started. Initial states with an entropy that is smaller than that of equilibrium occur for instance when milk is poured into coffee or when the dividing barrier between two different gases is removed. By going backwards in time, starting from equilibrium, the entropy of these systems would decrease, and the time evolution would tend to a highly ``improbable'' state.  We conclude that a time evolution towards a state of lower entropy (now going again forward in time) would in no way be in contradiction with classical mechanics. For such a case, one would have to conclude from a deterministic point of view that the initial state, even though we cannot resolve it to the required high precision, was one of the ``special'' initial states that tend to low-entropy states at some later time. Such states lie dense in the energy shell even though they have a measure of zero. 

\subsection{An alternative view}

By abandoning the idea that a position in phase space has infinite precision, all the problems raised in the previous subsection are solved. In order to show this, we start now from the assumption that a point in phase space has limited precision, and explore its consequences. Because Hamilton's equations do not fix unequivocally the future time evolution, we have to combine the idea of finite precision of phase space points with a rule that specifies which type of time evolution out of the possible ones is taken by the system. This rule is the rule of equal probabilities for (or at least of a smooth probability distribution over) those evolutions that are compatible with the initial state. 

The assumption that points in phase space have limited precision is logically very satisfying for several reasons. First, as already mentioned, it creates the room for employing probabilistic rules, which are the basis of statistical mechanics. Second, a finite precision of phase space points is all that is ever needed for ``deriving'' statistical mechanics from classical mechanics. In particular when discussing the concept of quasi-ergodicity one always resorts to considering phase space with a certain resolution, given by the mesh size of the grid mentioned above. Third, a finite precision of phase space permits a system to reach equilibrium within a short time and to truly forget the past. With infinite precision, the time required to visit all cells of the phase space grid is proportional to the number of cells, and many orders of magnitude larger than the age of the universe. When the initial state and all future states  have only finite precision, which we can take to be identical to the cell size, the number of cells which the system can reach within a certain time is larger than 1. This means that the number of cells that can be reached from the initial state increases exponentially in time, leading to the conclusion that after a short time the system could be anywhere on the energy shell. Furthermore, a true equilibrium state should have no traces of the past. However, with infinite precision the state of a system would always be uniquely related to a predecessor state at a previous moment in time. With finite precision, the number of possibly predecessor cells increases exponentially with the length of time by which one looks back. When this time interval becomes large enough, the system could have been anywhere on the energy shell, and the initial state is completely forgotten. 

All these considerations can be made based on classical mechanics alone in the attempt to reconcile it with the second law of thermodynamics. Of course, it is very satisfying to know that quantum mechanics confirms the suggestion that points in phase space have only limited precision. The uncertainty relation makes it impossible to fix simultaneously the momenta and positions of particles to infinite precision. Furthermore, quantum mechanics fixes the ``mesh size'' for the phase space grid. A cell in phase space has the size $\hbar ^{3N}$. 

An important conclusion from these considerations is that the time evolution of a thermodynamic systems is underdetermined by classical mechanics if the idea of infinite precision is abandoned. An additional law is required that specifies which type of time evolution is taken. This law is the law of equal probabilities, leading naturally to the second law of thermodynamics. Thus, the second law of thermodynamics is an emergent law in the strong sense; it is not contained in the microscopic laws of classical mechanics. 

\subsection{Other ways leading from classical mechanics to  the second law of thermodynamics}

There exist other approaches to statistical mechanics, which make expressly additional assumptions that are not part of classical mechanics. In their textbook on statistical physics, Landau and Lifschitz reject the idea of quasi-ergodicity of a large system because of the impossibility that a system visits even  a small part of phase space within the duration of an experiment. Instead, they divide the system in many small subsystems and note that observables are sum variables over all these subsystems. By assuming that these subsystems are statistically independent, equal a priori probabilities and the second law of thermodynamics can be obtained. By assuming statistical independence of the subsystems they make an asumption that is similar in spirit to the idea of the previous subsections. They assume that all future time evolutions that are compatible with the initial state are equally probable: Statistical independence of subystems means that the influence of one subsystem on a neighboring one does not depend on the specific microscopic state of the subsystem. Rather, the influence on the neighbor is a ``typical'' influence, as if the neighbor was in a random state. Specific correlations between subsystems or processes that have taken place in the past are irrelevant for equilibrium behavior. 

A similar type of assumption underlies Boltzmann's equation. 
This equation describes the time evolution of a gas of particles, which goes towards an equilibrium state. Since this equation appears very plausible, one is tempted to forget that it contains assumptions that are not part of classical mechanics. In particular, this equation relies on the assumption that correlations due to past processes are irrelevant. Bolzmann's equation can be derived from Hamilton's equations by making a few simplifications. It is based on the density of particles 
 $f(\vec p,\vec q,t)$ in 6-dimensional phase space. This phase space is the phase space of one particle, and the state of the system is represented by $N$ points in this phase space. Time evolution is determined by collisions between particles and by free motion in between. If no external potential is included, Boltzmann's equation reads
\begin{equation} \frac{\partial f(\vec p,\vec q,t)}{\partial t} + \dot {\vec
  q}\cdot \frac{\partial f(\vec p,\vec q,t)}{\partial \vec q} = \int d^3 p_2  \int d^3 p_3 \int d^3 p_4 
W(\vec p, \vec p_2; \vec p_3, \vec p_4)\left[f(\vec p_3,\vec q,t)f(\vec p_4,\vec
q,t)-f(\vec p,\vec q,t)f(\vec p_2,\vec q,t)\right]\, .
\label{boltzmann}
\end{equation}
Collisions between particles with momenta $\vec p_3$ and $\vec p_4$ lead to the momenta $\vec p$ and $\vec p_2$, or vice versa. The function $W$ contains the cross section for such collisions. The collision term depends only on the products of one-particle densities $f$, which means that the probabilities for particles being at the position $\vec q$ are assumed to be independent from each other. Correlations between particles, which are created by collisions, are thus neglected. The success of Boltzmann's equation justifies this assumption, and it means that a detailed memory of past processes is not required for predicting correctly the future time evolution. Again, we find that the future time evolution is assumed to be a ``typical'' time evolution, which results when the present microstate is a random state compatible with the observables, i.e., with the function $f(\vec p,\vec q)$. It is well known that Boltzmann's equation leads to a decrease in time of the so-called H-function, which is the integral of $(f \log f)$ over phase space. The system approaches equilibrium, where the H-function has its minimum, which is equivalent to the entropy having its maximum.

\section{From quantum mechanics to statistical mechanics}

There are essentially four approaches to connect quantum mechanics to statistical mechanics: The first approach consists in considering a $N$-particle wave function of a closed system. The second approach includes the interaction with an environment via a potential. The third approach models the environment as consisting of many degrees of freedom. The fourth approach treats quantum mechanics as an ensemble theory and views statistical mechanics as being part of quantum mechanics. We will briefly discuss in the following all four approaches, their achievements and their shortcomings. We shall again see that in the first three approaches additional assumptions must be made about correlations being absent and time evolutions showing ``typical'' behavior. 

\subsection{The eigenstate thermalization hypothesis}

A system of $N$ interacting particles in a potential well has chaotic dynamics in classical mechanics. In quantum mechanics, its eigenstate wave functions look very random. When dividing the volume of the potential well in to small subvolumes, we can expect that an eigenfunction of the hamiltonian has equal particle density in all subvolumes. The eigenstate thermalization hypothesis suggests that in fact all expectation values of observables, evaluated with an eigenfunction, correspond to the thermal average in equilibrium (Deutsch, 1991; Srednicki, 1994; Rigol et al, 2008). By using an appropriate superposition of eigenstates, one can generate an initial state that is far from equilibrium, and assuming that the expansion coefficients are such that they do not lead to special superpositions at later times, the system is bound to evolve towards an ``equilibrium'' state, where the observables have their according values. 

The problems with this approach are twofold: 

First, in order to obtain the ``typical'' time evolution towards equilibrium, we must make the assumption that there are no correlations in the expansion coefficients that might lead to special, low-entropy states at later times. This is the same type of assumption that is made in the approaches to classical mechanics that we discussed in the previous section. Second, this approach does not give the kind of density matrix that represents a thermodynamic equilibrium. The density matrix of this system is that of a pure state, with diagonal elements that are constant in time and that correspond to the weights of the different eigenfunctions in the initial state. In contrast, the density matrix at equilibrium is that of a mixed state, which can be represented as a diagonal matrix in the basis of energy eigenstates, with all entries being identical.

In the previous section, we have argued that the problems that arise when reconciling classical mechanics with the second law of thermodynamics are resolved when abandoning the idea of infinite precision. It appears to me that the quantum mechanical time evolution of a closed system can be reconciled with statistical mechanics in a similar way, by abandoning the idea of infinite precision of a wave function. Only if a wave function has finite precision does the statement make sense that the expansion coefficients have the ``most likely'' property of leading to no special macrostates at later times. Furthermore, if a  wave function has finite precision, it can be compatible with many pure and mixed (infinite-precision) states, i.e., with many different density matrices, and with many possible future time evolutions. Which time evolution is actually taken by the system has then to be fixed by additional laws.

\subsection{Interaction with the environment through a potential}

This approach is due to Felix Bloch (1989), and it starts by taking the Hamilton  operator to be that of an isolated system, $H_0$, plus an interaction potential that represents the effect of the environment. Since no system can be completely isolated from the rest of the world, there is always some external influence. 
Starting from an initial state $|\Psi,0\rangle = \sum_n
c_n |n\rangle$, the wave function becomes at later times
\be |\Psi,t\rangle = \sum_n c_n(t) e^{-iE_nt/\hbar}
|n\rangle \, . \label{psiwall} \ee 
Assuming that the interaction with the environment does (almost) not change the energy of the system, the energy eigenvalues $E_n$ are all within a very small interval. The time evolution of the coefficients $c_n$ can be calculated to be
\be i\hbar \frac{\partial
  c_m}{\partial t} =\sum_nV_{mn}(t)c_n(t) \ee with \be V_{nm}(t) =
\langle m | V | n\rangle \, e^{-i(E_n-E_m)t/\hbar} \, .  \ee 
Because the vector $c_n$ is normalized to 1 ($\sum_n |c_n|^2=1$), its tip moves on the unit sphere. If the system is ergodic in a suitable sense, the tip of the vector will come arbitrarily close to every point on this sphere. There is a very close analogy between the trajectory  $c_n(t)$ and the trajectory in phase space of a classical mechanics trajectory. All the reasoning made above in context of classical trajectories therefore applies also to $c_n(t)$. By abandoning the idea of infinite precision of a quantum mechanical state, the problem that egodicity requires incredibly long time periods would be resolved as well as the problem that a completely deterministic time evolution leaves no room for appeals to the ``most likely'' behavior.

\subsection{Coupling to an environment with many degrees of freedom}

Since the 1980s, a series of very fruitful investigations have been performed on the time evolution of a quantum mechanical system that interacts with an environment that consists of many degrees of freedom, all of which are also modelled quantum mechanically, for instance as harmonic oscillators. The time evolution of the system and all environmental degrees of freedom is taken to be a unitary time evolution according to the Schr\"odinger equation. Accordingly, the density matrix of the full system, which includes the environmental degress of freedom, is that of a pure state.  However, when focussing on the system of interest, the trace over the environmental degrees of freedom is taken, leading to a reduced density matrix, which is generally that of a mixed state. 

One important application of this procedure is quantum diffusion (Caldeira \& Leggett, 1983), where a particle is coupled to the environmental degrees of freedom via its position. Using the path integral formalism and taking in the end the limit $\hbar \to 0$, one finds that that probability density that the particle has moved during time $t$ by a distance $\vec r$ is described by the Fokker-Planck equation. Thus, the transition from a quantum mechanical description to a classical, stochastic description has been made. This type of phenomenon is called ``decoherence''.

Another important application of decoherence theory is that of quantum measurement. In this case, the system couples to the many degrees of freedom of the (macroscopic) measurement apparatus via the observable that is being measured, for instance the spin. In turns out that the reduced density matrix of the system becomes diagonal after a very short time (Zurek, 1991; Zeh, 2002). This means that it describes a classical probabilistic superposition of the different measurement outcomes. In a similar vein, one can argue that the density matrix of a thermodynamic system (or a small subsystem of it) will evolve to the density matrix of a classical superposition when the system is coupled to an environment that consists of many degrees of freedom, and when the trace over these degrees of freedom is taken. 

A close look at these calculations reveals that they rely on two types of assumptions that are similar in spirit to the assumptions that we have discussed before. First, some type of statistical independence or lack of special correlations between the variables that describe the environmental degrees of freedom must be assumed. The reduced density matrix of the system becomes diagonal only if the non-diagonal elements, which contain products of a function of the amplitudes and phases of the different environmental degrees of freedom, decrease to zero. Second, it must be assumed that for all practical purposes the entanglement of the system with the environment, which is pushed out of sight by taking the trace over the environmental degrees of freedom, can be ignored. Such an entanglement would contain a full memory of the process that has taken place since the system started to interact with the environment. Now, we can argue once more that a finite precision of the wave function would solve these conceptual problems: the entanglement with the environment and the perfect memory of the past could vanish with time, and the lack of special correlations could be phrased in terms of the most probable or typical time evolution. 

Critics of decoherence theory focus on its incomplete potential to explain the measurement process. If many experiments of the same type are performed, the diagonal entries of the density matrix tell correctly which proportion of experiments will show which measurement result. However, quantum mechanics is taken to be a theory that describes individal systems, not just ensembles of systems. This is not the topic of this article, but it leads us to the fourth approach to the relation between quantum and statistical mechanics.

\subsection{Quantum mechanics as a statistical theory that includes statistical mechanics}

A view of quantum mechanics that circumvents the need to reconcile the deterministic, time-reversible evolution of the Schr\"odinger equation  with the stochastic, irreversible features of statistical mechanics, is the statistical interpretation (Ballentine, 1970). In this interpretation, quantum mechanics is viewed as an ensemble theory, which gives probabilities for measurement outcomes. By extending this theory to include mixed states, statistical mechanics becomes part of quantum mechanics. The price for this elegant solution of our problem is the incompleteness of the theory, because individual systems cannot be described by it. However, this is again not the topic of this article. 

\section{Conclusions}

By taking a close look at the different ``derivations'' of statistical mechanics, in particular of the second law of thermodynamics, from classical or quantum mechanics, we have seen that all these derivations make similar assumptions that go beyond the deterministic, time-reversible microscopic theory from which they start. All derivations assume that the time evolution of the system is ``typical'' in some sense. This means that the most likely type of time evolution of observable variables, given our knowledge of the system, does occur. If the time evolution of the system was deterministic on the microscopic level, our probabilistic statements about the system would merely be due to our ignorance of the precise microscopic state. However, in this case there would be no conclusive reason why the time evolution of the system should comply with our ignorance and take the ``most likely'' route, because other routes would also be compatible with our knowledge of the initial state. Furthermore, when looking into the past, the evolution backwards in time does not take the most likely route. We must therefore understand the rule of equal probabilities as an additional law that is required to correctly describe the  behavior of the system. However, there would be no room for an additional law if the laws of classical or quantum mechanics did fully determine the time evolution of the system. We are therefore led to conclude that the description in terms of classical or quantum mechanics is only an approximate description, and that points in phase space or wave function have only a limited precision. This line of reasoning is consistent with other philosophical and scientific arguments. From the philosophical point of view, the concept of infinite precision of a state is a metaphysical idea that can not, even in principle, be tested. From a scientific point of view, we know that Newton's (or Hamilton's) equations of motion and the Schr\"odinger equation are only an approximation to reality. The fact that these two theories  work so well for many applications can make us blind to the the possibility that their limited precision may have strong effects in systems that consist of macroscopic numbers of nonlinearly interacting particles. Such complex systems are therefore not simply the sum of their parts, but are ruled by new laws that are not contained in a microscopic description.

\subsection*{References}

Ballentine, L.E. (1970). ``The statistical interpretation of quantum mechanics.'' in \emph{Reviews of Modern Physics} 42, 358-381.

\medskip
\noindent
Beisbart, C., Hartmann, S., eds (2011) \emph{Probabilites in Physics} Oxford University Press. 

\medskip

\noindent
Bloch, F. (1989) \emph{Fundamentals of Statistical Mechanics. Manuscript and Notes of Felix Bloch.} Edited by John Dirk Walecka. Stanford University Press 1989, World Scientific 2000.

\medskip

\noindent
Caldeira, A.O., Leggett, A.J. (1983) ``Path integral approach to quantum Brownian motion.'' \emph{Physica A} 121, 587-616.

\medskip

\noindent
Deutsch, J. M. (1991). ``Quantum statistical mechanics in a closed system.'' \emph{Phys. Rev. A} 43, 2046-2049.

\medskip

\noindent
Kohn, W. (1999). ``Electronic structure of matter-wave functions
and density functionals.'' \emph{Reviews of Modern Physics} 71, 1253-1266.

\medskip

\noindent
Rigol, M., Dunjko, V., Olshanii, M. (2008). ``Thermalization and its mechanism for generic isolated quantum systems.'' \emph{Nature} 452, 854-858.

\medskip

\noindent
Schlosshauer, M. (2004) ``Decoherence, the measurement problem, and interpretations of quantum mechanics.'' \emph{Reviews of Modern Physics} 76, 1268-1305.

\medskip

\noindent
Srednicki, M. (1994). ``Chaos and quantum thermalization.'' \emph{Phys. Rev. E} 50, 888-901.

\medskip

\noindent
Zeh, H.D. (2002) ``The wave function: It or bit?'' \emph{ arXiv:quant-ph/0204088v2}

\medskip

\noindent
Zurek, W.H. (1991) ``Decoherence and the transition from quantum to classical.'' \emph{Physics Today} 44, 36-44.

\end{document}